# STATISTICAL MODEL OF DOWNLINK POWER CONSUMPTION IN CELLULAR CDMA NETWORKS

Stylianos P. Savaidis[1] and Nikolaos I. Miridakis [2, 3]

[1]*Department of Electronics and* [2]*Department of Computer Engineering, Technological Educational Institute (TEI) of Piraeus, 250 Thivon & P.Ralli, Aigaleo, Athens–12244, Greece*

[3]*Department of Informatics, University of Piraeus, 80 Karaoli & Dimitriou, 185 34 Piraeus, Greece*

**Abstract**—Present work proposes a theoretical statistical model of the downlink power consumption in cellular CDMA networks. The proposed model employs a simple but popular propagation model, which breaks down path losses into a distance dependent and a log-normal shadowing loss term. Based on the aforementioned path loss formalism, closed-form expressions for the first and the second moment of power consumption are obtained taking into account conditions placed by cell selection and handoff algorithms. Numerical results for various radio propagation environments and cell selection as well as handoff schemes are provided and discussed.

## I. INTRODUCTION

Code division multiple access (CDMA) have been adopted by narrowband 2G and wideband 3G cellular wireless networks, due to its inherent virtue of providing a single frequency reuse pattern. Since the available spectrum is shared among all active users, the transmission power is the basic radio resource of CDMA based systems. In this context, power consumption becomes the dominant performance evaluation figure that determines network resource allocation and capacity.

Power consumption depends on the location of the mobile station (*MS*), traffic parameters and the QoS requirements of each service, experienced interference level as well as cell selection and handoff settings. Thus, the development of a power consumption model which takes into account the aforementioned parameters is a prerequisite for efficient deployment of CDMA networks. Typically, research activities on the area can be classified into those that examine the uplink [1]-[8] and the ones referring to the downlink direction [1], [3], [9]-[18]. Taking into account the asymmetric nature of data flows, the downlink is most likely to be the bottleneck point of CDMA networks. In addition, research studies of the uplink have provided analytical methodologies concluding to closed form expressions [19], which can tackle both hard and soft handoff connection modes. Typically, the downlink studies conclude to numerical simulations [1], [13], [14], [19], assumptions that simplify the examined network scenarios [3], [9], [12] or approximations that mainly resolve the complexity of calculations regarding soft handoff connection modes [9]-[11], [15]-[18]. Thus, modeling of the downlink in CDMA cellular networks is a rather important but laborious task.

Several research studies, as mentioned before, have developed an analytical methodology for the downlink performance evaluation but they resort to Monte Carlo simulations, when soft handoff is taken into account [1], [13], [14], [19]. In [3], [10] and [12] an analytical framework with closed-form expressions has been obtained but these works do not consider the soft handoff option, which requires particular attention in CDMA networks. Both hard and soft handoff connection modes are analyzed in [9] but the obtained closed-form expressions estimate the minimum downlink capacity. In [11] the complicated sums of the log-normal interferences that typically appear in soft handoff connection mode have been approximated by a log-normal distribution, which concludes to closed form expressions regarding the downlink capacity. Apart from the aforementioned approximation, the capacity evaluation in [11] simplifies the impact of soft handoff assuming that interference contributed by the soft handoff users is double, when compared with the hard handoff users. In [15] a rather efficient calculation methodology is introduced, which can estimate downlink capacity and outage probability considering both Active Set (*AS*) size and soft handoff option. The proposed methodology provides general analytical expressions but it demonstrates a rather high computational load, whereas the capacity calculations are possible using approximations according to the Central Limit Theory. A soft handoff scheme aiming to minimize power consumption and increase connection reliability is introduced in [16]. The proposed model in [16] approximates the sums of log-normally distributed random variables appearing in the various expressions as a single log-normal variable. Closed form expressions for the average power consumption are provided in [17] but still the numerical implementation requires a Monte Carlo simulation under soft handoff conditions and balanced power allocation for the involved Base Stations (*BS*s). In [18] an alternative calculation methodology is introduced in order to derive closed form expressions of the capacity at a certain outage probability. Nevertheless, the former expressions were obtained using an approximation of the energy per bit to interference ratio introducing a "macrodiversity non-orthogonality factor" and Gamma approximations of the interferences and signals in soft handoff conditions.

According to the above mentioned description, the development of a theoretical statistical model that facilitates performance evaluation of the downlink in cellular CDMA networks becomes quite laborious, especially when soft handoff is considered. Approximating assumptions or numerical simulations are typically employed in order to overcome the complexity of analysis. The presence of sums of log-normally distributed random variables in the various expressions is the major obstacle regarding the derivation of closed-form analytical expressions. The present work proposes an alternative approach in order to overcome this kind of complexity and conclude to closed form expressions. In particular, a Taylor series expansion of the aforementioned complicated expressions is employed, which next makes possible a straightforward calculation of power consumption moments. In fact present work demonstrates the calculation procedure for the first two moments of power consumption, although in principle affords calculation of higher order moments. The proposed calculation scheme can integrate several realistic conditions including a best BS selection condition, the impact of a soft handoff threshold as well as *AS* size. Overall, the present work provides a theoretical statistical model, which attempts to balance efficiently between the assumptions that oversimplify the examined network scenarios, the inaccuracies of the potential approximations and the physical insight that a closed form expression may provide.

Section II, describes the radio propagation model and the downlink power consumption formulas for hard handoff (*HHO*), 2-way and 3-way soft handoff (*SHO*) connection modes. Section III describes the conditions placed by cell selection and handoff schemes. In Section IV, the calculation details for the first and second moments of the downlink power consumption are discussed in details. Section V includes numerical results and verification tests regarding the proposed calculation scheme. Finally, section VI summarizes the main conclusions and discusses potential extensions of current work.

## II. DOWNLINK POWER CONSUMPTION

The adopted radio propagation model assumes that fast fading can be compensated by special reception techniques, e.g. rake receiver, thus it can be considered as a pure large scale path loss model. In particular, path losses are solely determined by a path loss factor, which determines the distance based losses, and a shadowing loss component, which demonstrates a log-normal behavior. Thus, the power received from a transmitting BS can be determined by the following expression:

$$P(r,\zeta) = r^{-\alpha} 10^{\zeta/10} P_T \qquad (1)$$

where $r$ denotes the distance between MS and BS and $P_T$ the BS's total transmitted power; $\alpha$ is the path loss factor and $\zeta$ denotes the shadowing losses as a zero-mean Gaussian distributed random variable with standard deviation $\sigma$. The shadowing loss random variable for a certain BS, i.e. $BS_i$, can be further analyzed into two components, namely $\zeta_i = a\xi + b\xi_i$ [13], [14]. The $a\xi$ component denotes a part of shadowing that is common for all BSs and it represents the environment near and around the MS, whereas $b\xi_i$ denotes shadowing effects that depend on the environment near and around BS. The constants $a$ and $b$, fulfill the relationship $a^2+b^2=1$, whereas $\xi_i$ are considered as independent zero-mean Gaussian distributed random variables with standard deviation $\sigma$ [12]-[14].

The network scenario under investigation considers center feed cells of hexagonal shape and equal size. The interference and downlink power consumption analysis assumes an MS, which camps in cell 1 with two tiers of neighboring cells around it, as Fig. 1 depicts. Intra-cell interference calculations require only the knowledge of the distance $r_1$ between the serving $BS_1$ and the MS. However, for inter-cell interference calculations, both distance $r_1$ and angular position $\theta_1$, as Fig. 2 shows, should be considered. The distance $r_1$ between MS and $BS_1$ varies from zero to $\sqrt{3}R\cos(\theta_1)/2$, whereas the angular coordinate $\theta_1$ varies from 0° to 360°. Due to the hexagonal symmetry, throughout the remaining analysis only angular positions $\theta_1=0° \sim 30°$, will be examined.

Power control function should under ideal conditions regulate downlink power consumption in order to lock energy per bit to interference value to the target value $[E_b/I_o]_t$ required by each service. Thus, by calculating interference level and assuming a perfect power control scheme, downlink power consumption for HHO, 2-way and 3-way SHO connection modes can be estimated as follows:

*A) Hard Handoff Scenario*

When a HHO connection mode is assumed, all downlink transmissions to other MSs within the cell as well as in neighbor cells are considered interference. In principle, the proposed model can tackle network scenarios with unequal traffic loads per cell and thus different transmit power level $P_{T,i} = \delta_i P_T$ per base station. However, in order to simplify model's analysis, we assume equal total downlink transmission levels $P_T$ in each cell (i.e. $\delta_i = 1$). In this respect, power consumption for a single connection in cell 1 can be calculated as follows [13]-[14]:

$$\left[\frac{E_b}{I_o}\right]_t = \frac{W}{vR} \frac{r_1^{-\alpha} 10^{\zeta_1/10} P_{s1}}{(1-u) r_1^{-\alpha} 10^{\zeta_1/10} (P_T - P_{s1}) + \sum_{i=2}^{19} r_i^{-\alpha} 10^{\zeta_i/10} P} \Rightarrow$$

$$\Rightarrow P_{s1} = C_t X(\underline{\xi}) P_T = C_t \left[ \sum_{i=0, i \neq 1}^{19} X_i \right] P_T = \beta_1 P_T \qquad (2)$$

where $C_t = vR[E_b/I_o]_t/W$, $X_0 = 1 - u$ denotes intra-cell interference, $X_i = C_{1,i} 10^{b(\xi_i - \xi_1)/10}$ denotes inter-cell interference and $C_{1,i} = (r_1/r_i)^\alpha$. Vector $\underline{\xi} = (\xi_1, ..., \xi_{19})$ denotes the uncorrelated shadowing random variables of $BS_i$s, $v$ is the activity factor which applies to the service under examination, $R$ is the service data rate, $W$ is the chip rate, $u$ denotes the orthogonality between the various transmissions and $\beta_1$ is the fraction of $P_T$ allocated for a single link. For the sake of simplicity, in equation (2) and throughout equations (3) and (4), we assume that $(P_T - P_{s1}) \cong P_T$ as far as it concerns intracell interference calculations.

*B) 2-way Soft Handoff Scenario*

If we assume a maximal ratio combination capability (MRC) and a balanced power allocation scheme ($P_{s1} = P_{sk} = P_{s,1k}$) among *BS*s in cell 1 and cell *k*, which participate in the 2-way SHO connection, then power consumption for a single connection in cell 1 is calculated as follows [9], [13]-[14]:

$$\left[\frac{E_b}{I_o}\right]_t = \left[\frac{E_b}{I_o}\right]_1 + \left[\frac{E_b}{I_o}\right]_k = \frac{W}{vR} \left[ \frac{r_1^{-\alpha} 10^{\zeta_1/10} P_{s1}}{(1-u) r_1^{-\alpha} 10^{\zeta_1/10} P_T + \sum_{i=2}^{19} r_i^{-\alpha} 10^{\zeta_i/10} P_T} + \right.$$

$$\left. + \frac{r_k^{-\alpha} 10^{\zeta_k/10} P_{sk}}{(1-u) r_k^{-\alpha} 10^{\zeta_k/10} P_T + \sum_{i=1, i \neq k}^{19} r_i^{-\alpha} 10^{\zeta_i/10} P_T} \right] \Rightarrow P_{s,1k} = C_t \left[ \frac{1}{X(\underline{\xi})} + \frac{1}{Y(\underline{\xi})} \right] P_T =$$

$$= C_t \left[ \left( \sum_{i=0, i \neq 1}^{19} X_i \right)^{-1} + \left( \sum_{i=0, i \neq k}^{19} Y_i \right)^{-1} \right] P_T = \beta_{1k} P_T \qquad (3)$$

where similar to eq. (2), $Y_0 = 1 - u$, $Y_i = C_{k,i} 10^{b(\xi_i - \xi_k)/10}$ and $C_{k,i} = (r_k/r_i)^\alpha$; $\beta_{1k}$ is the fraction of $P_T$ allocated by each *BS*, which participates to the 2-way *SHO* connection.

*C) 3-way Soft Handoff Scenario*

If we assume MRC reception conditions and balanced power allocation scheme ($P_{s1} = P_{sk} = P_{sl} = P_{s,1kl}$) between *BS*s in cell 1, cell *k* and cell *l*, which participate in the 3-way SHO connection, then power consumption for a single connection in cell 1 is calculated as in the previous cases [9], [13]-[14]:

$$\left[\frac{E_b}{I_o}\right]_t = \left[\frac{E_b}{I_o}\right]_1 + \left[\frac{E_b}{I_o}\right]_k + \left[\frac{E_b}{I_o}\right]_l = \frac{W}{vR}\left[\frac{r_1^{-\alpha}10^{\zeta_1/10}P_{s1}}{(1-u)r_1^{-\alpha}10^{\zeta_1/10}P_T + \sum_{i=2}^{19}r_i^{-\alpha}10^{\zeta_i/10}P_T} + \right.$$

$$\left. + \frac{r_k^{-\alpha}10^{\zeta_k/10}P_{sk}}{(1-u)r_k^{-\alpha}10^{\zeta_k/10}P_T + \sum_{i=1,i\neq k}^{19}r_i^{-\alpha}10^{\zeta_i/10}P_T} + \frac{r_l^{-\alpha}10^{\zeta_l/10}P_{sl}}{(1-u)r_l^{-\alpha}10^{\zeta_l/10}P_T + \sum_{i=1,i\neq l}^{19}r_i^{-\alpha}10^{\zeta_i/10}P_T}\right] \Rightarrow$$

$$\Rightarrow P_{s,1kl} = C_t\left[\frac{1}{X(\underline{\xi})} + \frac{1}{Y(\underline{\xi})} + \frac{1}{Z(\underline{\xi})}\right]P_T = C_t\left[\left(\sum_{i=0,i\neq 1}^{19}X_i\right)^{-1} + \left(\sum_{i=0,i\neq k}^{19}Y_i\right)^{-1} + \left(\sum_{i=1,i\neq l}^{19}Z_i\right)^{-1}\right]P_T = \beta_{1kl}P_T$$

(4)

where similar to eq. (2)-(3), $Z_0 = 1-u$, $Z_i = C_{l,i}10^{b(\xi_i-\xi_l)/10}$ and $C_{l,i} = (r_l/r_i)^{\alpha}$. $\beta_{1kl}$ is the fraction of $P_T$ allocated by each BS, which participates to the 3-way *SHO* connection.

At this point it should be mentioned that in principle, the proposed model can tackle both balanced and unbalanced power allocation schemes by defining different weights on $P_{s1}, P_{sk}$ and $P_{sl}$. Nevertheless, for simplicity reasons, in our analysis we assume equal weights on $P_{s1}$ $P_{sk}$ and $P_{sl}$, yet without loss of generality.

## III. CELL SELECTION AND HANDOFF SCHEMES

Cell selection and handover schemes influence the network performance [20], [21] and thus current section will examine the conditions that are imposed in our calculations by the aforementioned schemes. If cell 1 is the camping cell and assuming a best BS selection condition, then the transmission of cell-1 will be the best among the candidate cells *i* (=2, 3,…, 19), i.e. $\xi_i \leq \xi_1$-$R_{1,i}$ ($R_{1,i} = 10\log(C_{1,i})/b$). The former condition describes an ideal cell selection scenario and a perfect power control scheme. The addition of a hysteresis threshold *cst*=10log(*CST*)/*b* can account for possible cell selection and power control imperfections e.g. $\xi_i \leq \xi_1$-$R_{1,i}$+*cst*.

Apart from the above described conditions, the handoff algorithm is placing additional ones. The handoff scheme considered here is one that accepts a maximum number of simultaneous physical connections equal to the *AS* size. In addition, the algorithm places a *SHO* threshold in order to accept a *BS* to join the *AS*. If *SHO* is not an option, i.e. *AS*=1, then the handoff condition is identical to the cell selection one. However, if *AS*>1, then the *HHO* scenario implies that the signal strength of all monitored *BS*s should not exceed the *SHO* threshold. The latter statement is expressed as $\xi_i \leq \xi_1$-$R_{1,i}$-*sht*, (*sht*=10log(*SHT*)/*b*). Concluding with the *HHO* mode, the following conditions apply:

$$\xi_i \leq \xi_1 - R_{1,i} - sht, \; AS > 1 \quad (5)$$

$$\xi_i \leq \xi_1 - R_{1,i} + cst, \; AS = 1 \quad (6)$$

If 2-way *SHO* conditions apply, then two simultaneous connections with $BS_1$ and $BS_k$ occur. If *AS*>1, then $BS_k$'s signal is the strongest signal among the monitored ones and exceeds *SHO* threshold. After some straightforward calculations, the former statements can be described as follows:

$$\xi_1 - R_{1,k} - sht \leq \xi_k \leq \xi_1 - R_{1,k} + cst, \ k \in AS > 1 \quad (7)$$

$$\xi_i \leq \xi_k - R_{k,i}, \ i \notin AS(=2) \quad (8)$$

$$\xi_i \leq \xi_1 - R_{1,i} - sht, \ i \notin AS(=3) \quad (9)$$

Finally, when *AS*=3 a 3-way *SHO* scenario applies and a single logical network link include physical links with three *BS*s, e.g. $BS_1$, $BS_k$ and $BS_l$. $BS_k$'s and $BS_l$'s signal are the strongest signals among the monitored ones and both exceed the *SHO* threshold. Assuming that $BS_l$'s signal is the weakest among the *AS* participants, then all other monitored signals should be weaker than $BS_l$'s signal. After some straightforward calculations the former statements can be expressed as follows:

$$\xi_1 - R_{1,k(l)} - sht \leq \xi_{k(l)} \leq \xi_1 - R_{1,k(l)} + cst, \ k(l) \in AS \quad (10)$$

$$\xi_l \leq \xi_k - R_{k,l}, \ k \ and \ l \in AS \quad (11)$$

$$\xi_i \leq \xi_l - R_{l,i}, \ i \notin AS \quad (12)$$

Concluding, it is worthwhile to mention that no restrictions are placed in non monitored cells, which *de facto* do not participate to handoff process. In order to simplify the analysis throughout the remaining analysis all cells in both tiers will be considered as monitored.

## IV. DOWNLINK POWER CONSUMPTION STATISTICS

Three handoff schemes are considered in this section, i.e. *AS*=1, 2 and 3. In all following calculations, the random shadowing loss values $\xi_i$ are restricted by the cell selection and handoff conditions discussed in the previous section. If *AS*=*m* and $\Omega^m$ is the subset of random $\xi_i$ values, which allow *MS* to camp in cell 1, then $\Omega^m$ can be expressed as $\Omega^m = \Omega_1^m \cup \sum_k \Omega_{1k}^m \cup \sum_{k,l} \Omega_{1kl}^m$. Subsets $\Omega_1^m$, $\Omega_{1k}^m$ and $\Omega_{1kl}^m$ include all $\xi_i$ values, which conform to *HHO*, 2-way *SHO* and 3-way *SHO* conditions, respectively. The conditions for each subset are established with eqs. (5)-(6), (7)-(9) and (10)-(12) of section III. Apparently $\Omega_{1k}^1 = \varnothing$ and $\Omega_{1kl}^1 = \Omega_{1kl}^2 = \varnothing$.

The above discussed subsets correspond to all possible connection modes that may occur in the cell under investigation i.e. cell 1. If the downlink transmitted power for a single user in cell 1 is $P_s = \beta P_T$, then the actual point of interest in our calculations is the fraction *β* of the total transmitted power. The first and the second moment of *β* can be obtained as

$$E[\beta | \underline{\xi} \in \Omega^m] = \frac{P(\Omega_1^m)E[\beta_1 | \underline{\xi} \in \Omega_1^m] + \sum_k P(\Omega_{1k}^m)E[\beta_{1k} | \underline{\xi} \in \Omega_{1k}^m] + \sum_{k,l} P(\Omega_{1kl}^m)E[\beta_{1kl} | \underline{\xi} \in \Omega_{1kl}^m]}{P(\Omega^m)} \quad (13)$$

$$E[\beta^2 | \underline{\xi} \in \Omega^m] = \frac{P(\Omega_1^m)E[\beta_1^2 | \underline{\xi} \in \Omega_1^m] + \sum_k P(\Omega_{1k}^m)E[\beta_{1k}^2 | \underline{\xi} \in \Omega_{1k}^m] + \sum_{k,l} P(\Omega_{1kl}^m)E[\beta_{1kl}^2 | \underline{\xi} \in \Omega_{1kl}^m]}{P(\Omega^m)} \quad (14)$$

where $P(\Omega^m) = P(\Omega_1^m) + \sum_k P(\Omega_{1k}^m) + \sum_{k,l} P(\Omega_{1kl}^m)$.

Since the shadowing random variables $\xi_i$ are independent their joint pdf is

$$f_{\underline{\xi}}(\underline{\xi}) = \prod_{n=1}^{19} f_{\xi_i}(\xi_i) = \prod_{n=1}^{19} \frac{e^{(\xi_i^2/2\sigma^2)}}{\sqrt{2\pi}\sigma} \qquad (15)$$

and thus $P(\Omega_1^m)$ can be calculated as follows:

$$P(\Omega_1^m) = \int_{-\infty}^{+\infty} f_{\xi_1}(\xi_1) \left\{ \prod_{n=2}^{19} \int_{-\infty}^{a_n(\xi_1)} f_{\xi_n}(\xi_n) \right\} d\xi_1 = \int_{-\infty}^{+\infty} f_{\xi_1}(\xi_1) \left\{ \prod_{n=2}^{19} A(a_n(\xi_1),0) \right\} d\xi_1 \qquad (16)$$

where we define function $A(x,y)$ as

$$A(x,y) = \exp\left[y^2\sigma^2 b^2 \ln(10)^2/200\right] \left\{ 0.5 + 0.5\,erf\left[\frac{x}{\sigma\sqrt{2}} - y\frac{\sigma b \ln(10)}{10\sqrt{2}}\right] \right\} \qquad (17)$$

and $a_n(\xi_1)$ is the upper limit of inequality (5) or (6), when $m>1$ or $m=1$, respectively. In a similar manner $P(\Omega_{1k}^m)$ is obtained by the following expressions:

$$P(\Omega_{1k}^m) = \int_{-\infty}^{+\infty} f_{\xi_1}(\xi_1) \left\{ \int_{b_k(\xi_1)}^{a_k(\xi_1)} f_{\xi_k}(\xi_k) \left[ \prod_{n=2, n\neq k}^{19} A(a_n(\xi_k),0) \right] d\xi_\kappa \right\} d\xi_1 \qquad (18)$$

where $a_k(\xi_1)$ and $b_k(\xi_1)$ is the upper and the lower limit of inequality (7), respectively. If $m=2$ then $a_n(\xi_k)$ is the upper limit of inequality (8), otherwise $a_n(\xi_k)(=a_n(\xi_1))$ is the upper limit of equation (9). In addition, if $m=2$ the integration over $\xi_k$ can be only evaluated numerically, whereas for $m=3$ the integration over $\xi_k$ is evaluated analytically as $[A(b_k(\xi_1),0)-A(a_k(\xi_1),0)]$. With a similar manipulation $P(\Omega_{1kl}^m)$ is obtained by the following expression:

$$P(\Omega_{1kl}^m) = \int_{-\infty}^{+\infty} f_{\xi_1}(\xi_1) \left\{ \int_{b_k(\xi_1)}^{a_k(\xi_1)} f_{\xi_k}(\xi_k) \left( \int_{b_l(\xi_k)}^{a_l(\xi_k)} f_{\xi_l}(\xi_l) \left[ \prod_{n=2, n\neq k,l}^{19} A_n(a_n(\xi_l),0) \right] d\xi_l \right) d\xi_k \right\} d\xi_1 \qquad (19)$$

where $a_k(\xi_1)$ and $b_k(\xi_1)$ is the upper and the lower limit of eq. (10), $a_l(\xi_k)$ and $b_l(\xi_1)$ is the upper and the lower limit of eqs. (11) and (10), respectively, whereas $a_n(\xi_l)$ is the upper limit of eq. (12).

*A) HHO Calculations*

According to equation (2) the first and the second moments of $\beta_1$ can be obtained as follows:

$$E[\beta_1] = C_t \sum_{\substack{i=0 \\ i\neq 1}}^{19} E[X_i], \quad E[\beta_1^2] = C_t^2 \left\{ \sum_{\substack{i=0 \\ i\neq 1}}^{19} E[X_i^2] + \sum_{\substack{j=0 \\ j\neq 1}}^{19} \sum_{\substack{i=0 \\ i\neq 1, i\neq j}}^{19} E[X_j X_i] \right\} \qquad (20)$$

where for $i\neq 0$

$$E[X_i] = \frac{C_{1,i}}{P(\Omega_1^m)} \int_{-\infty}^{+\infty} 10^{-b\xi_1/10} f_{\xi_1}(\xi_1) A(a_i(\xi_1),1) \Pi_i(\xi_1) d\xi_1 \quad (21)$$

$$E[X_i^2] = \frac{C_{1,i}^2}{P(\Omega_1^m)} \int_{-\infty}^{+\infty} 10^{-b\xi_1/5} f_{\xi_1}(\xi_1) A(a_i(\xi_1),2) \Pi_i(\xi_1) d\xi_1 \quad (22)$$

$$E[X_i X_j] = \frac{C_{1,i} C_{1,j}}{P(\Omega_1^m)} \int_{-\infty}^{+\infty} 10^{-b\xi_1/5} f_{\xi_1}(\xi_1) A(a_i(\xi_1),1) A(a_j(\xi_1),1) \Pi_{i,j}(\xi_1) d\xi_1 \quad (23)$$

and $\Pi_i(\xi_1) = \prod_{n=2, n \neq i}^{19} A(a_n(\xi_1), 0)$, $\Pi_{i,j}(\xi_1) = \prod_{n=2, n \neq i, j}^{19} A(a_n(\xi_1), 0)$. In addition, $E[X_0] = (1-u)/P(\Omega_1^m)$, $E[X_0^2] = (1-u)^2/P(\Omega_1^m)$, $E[X_0 X_j] = (1-u) E[X_j]$ with $E[X_j]$ given by (21) if we replace $j$ with $i$. The integration limits of the above expressions are the same with the ones appearing in eq. (16).

*B) 2-way SHO Calculations*

According to eq. (3) the first and the second moment of $\beta_{1,k}$, , can not be evaluated by employing the straightforward semi-analytical approach of subsection IV.A. In order to overcome this constraint, $\beta_{1k}$ is approximated by a Taylor expansion in the neighborhood of $E[X(\xi)]$ and $E[Y(\xi)]$. Next, by omitting Taylor series terms higher than the second order we conclude to (see Appendix I):

$$E[\beta_{1k} | \underline{\xi} \in \Omega_{1k}^m] = C_t \left\{ \frac{\overline{XY}}{\overline{X} + \overline{Y}} - \frac{\left[\overline{X^2} - (\overline{X})^2\right](\overline{Y})^2 + \left[\overline{Y^2} - (\overline{Y})^2\right](\overline{X})^2 - 2\left[\overline{XY} - \overline{X}\,\overline{Y}\right]\overline{XY}}{\left[\overline{X} + \overline{Y}\right]^3} \right\} \quad (24)$$

$$E[\beta_{1k}^2 | \underline{\xi} \in \Omega_{1k}^m] = C_t^2 \left\{ \frac{(\overline{Y})^4 \overline{X^2} + (\overline{X})^4 \overline{Y^2} + 2(\overline{XY})^2 \overline{XY}}{\left[\overline{X} + \overline{Y}\right]^4} + 2 \frac{\overline{XY}}{\overline{X} + \overline{Y}} \left[ \frac{E[\beta_{1k} | \underline{\xi} \in \Omega_{1k}^m]}{C_t} - \frac{\overline{XY}}{\overline{X} + \overline{Y}} \right] \right\} \quad (25)$$

where $\overline{X}$, $\overline{Y}$, $\overline{X^2}$, $\overline{Y^2}$ and $\overline{XY}$ correspond to $E[X | \underline{\xi} \in \Omega_{1k}^m]$, $E[Y | \underline{\xi} \in \Omega_{1k}^m]$, $E[X^2 | \underline{\xi} \in \Omega_{1k}^m]$, $E[Y^2 | \underline{\xi} \in \Omega_{1k}^m]$ and $E[XY | \underline{\xi} \in \Omega_{1k}^m]$, respectively. Following a calculation scheme as in section IV.A, the above mentioned E[.] terms can be expressed as a summation of all possible combinations of $E[X_i]$, $E[X_i^2]$, $E[X_i X_j]$, $E[Y_i]$, $E[Y_i^2]$, $E[Y_i Y_j]$ and $E[X_i Y_j]$. Also, each E[.] term can be expressed in an integral closed form expression, where the various integration limits are identical to the ones appearing in eq. (18). The $E[X_i]$ expression is obtained as follows:

$$E[X_i] = \frac{C_{1,i}}{P(\Omega_{1k}^m)} \int_{-\infty}^{+\infty} 10^{-\frac{b\xi_1}{10}} f_{\xi_1}(\xi_1) d\xi_1 \times \begin{cases} \int_{b_k(\xi_1)}^{a_k(\xi_1)} f_{\xi_k}(\xi_k) A(a_i(\xi_k),1) \Pi_{i,k}(\xi_k) d\xi_k, i \neq k \\ \int_{b_k(\xi_1)}^{a_k(\xi_1)} 10^{\frac{b\xi_k}{10}} f_{\xi_k}(\xi_k) \Pi_k(\xi_k) d\xi_k, i = k \end{cases} \quad (26)$$

If *m*=3 and *i*≠*k* (*i*=*k*) the *k*[th] integral in eq. (26) can be evaluated as [$A(b_k(\xi_1),0)-A(a_k(\xi_1),0)$] ([$A(b_k(\xi_1),1)-A(a_k(\xi_1),1)$]). The $E[Y_i]$ calculations are similar to eq. (26) with one difference, namely, the term $10^{-b\xi_1/10}$ is transferred to the *k*[th] integral as $10^{-b\xi_k/10}$

$$E[Y_i] = \frac{C_{k,i}}{P(\Omega_{1k}^m)} \times \begin{cases} \int_{-\infty}^{+\infty} f_{\xi_1}(\xi_1)d\xi_1 \int_{b_k(\xi_1)}^{a_k(\xi_1)} 10^{-\frac{b\xi_k}{10}} f_{\xi_k}(\xi_k) A(a_i(\xi_k),1) \Pi_{i,k}(\xi_k) d\xi_k, i \neq 1 \\ \int_{-\infty}^{+\infty} 10^{\frac{b\xi_1}{10}} f_{\xi_1}(\xi_1)d\xi_1 \int_{b_k(\xi_1)}^{a_k(\xi_1)} 10^{-\frac{b\xi_k}{10}} f_{\xi_k}(\xi_k) \Pi_k(\xi_k) d\xi_k, i = 1 \end{cases} \quad (27)$$

If *m*=3 the *k*[th] integral in eq. (27) can be evaluated as [$A(b_k(\xi_1),-1)-A(a_k(\xi_1),-1)$]. Similar to the *HHO* case $E[X_0] = E[Y_0] = (1-u)/P(\Omega_{1k}^m)$.

The $E[X_i^2]$ and $E[Y_i^2]$ expressions can be obtained from eqs. (26) and (27), respectively, if we substitute $10^{\pm b\xi_1/10}$ and $10^{\pm b\xi_k/10}$ with $10^{\pm b\xi_1/5}$ and $10^{\pm b\xi_k/5}$, respectively, $A(a_i(\xi_k),1)$ with $A(a_i(\xi_k),2)$ and [$A(b_k(\xi_1),\pm 1)-A(a_k(\xi_1),\pm 1)$] with [$A(b_k(\xi_1),\pm 2)-A(a_k(\xi_1),\pm 2)$]. Similar to the *HHO* case $E[X_0^2] = E[Y_0^2] = (1-u)^2/P(\Omega_{1k}^m)$.

The terms $E[X_iX_j]$ are described by the following integral expression:

$$E[X_iX_j] = \frac{C_{1,i}C_{1,j}}{P(\Omega_{1k}^m)} \int_{-\infty}^{+\infty} 10^{-\frac{b\xi_1}{5}} f_{\xi_1}(\xi_1)d\xi_1 \times \begin{cases} \int_{b_k(\xi_1)}^{a_k(\xi_1)} f_{\xi_k}(\xi_k) A(a_i(\xi_k),1) A(a_j(\xi_k),1) \Pi_{i,j,k}(\xi_k) d\xi_k, i,j \neq k \\ \int_{b_k(\xi_1)}^{a_k(\xi_1)} 10^{\frac{b\xi_k}{10}} f_{\xi_k}(\xi_k) A(a_{i(j)}(\xi_k),1) \Pi_{i,j,k}(\xi_k) d\xi_k, i(j) = k \end{cases} \quad (28)$$

where $\Pi_{i,j,k}(\xi_k) = \prod_{n=2, n \neq i,j,k}^{19} A(a_n(\xi_k),0)$. If *j*=3 and *i, j*≠*k* (*i* or *j*=*k*) the *k*[th] integral in eq. (28) can be evaluated as [$A(b_k(\xi_1),0)-A(a_k(\xi_1),0)$] ([$A(b_k(\xi_1),1)-A(a_k(\xi_1),1)$]).

The expression for the $E[Y_iY_j]$ term is given by the following equation:

$$E[Y_iY_j] = \frac{C_{k,i}C_{k,j}}{P(\Omega_{1k}^m)} \times \begin{cases} \int_{-\infty}^{+\infty} f_{\xi_1}(\xi_1)d\xi_1 \int_{b_k(\xi_1)}^{a_k(\xi_1)} 10^{-\frac{b\xi_k}{5}} f_{\xi_k}(\xi_k) A(a_i(\xi_k),1) A(a_j(\xi_k),1) \Pi_{i,j,k}(\xi_k) d\xi_k, i,j \neq 1 \\ \int_{-\infty}^{+\infty} 10^{\frac{b\xi_1}{10}} f_{\xi_1}(\xi_1)d\xi_1 \int_{b_k(\xi_1)}^{a_k(\xi_1)} 10^{-\frac{b\xi_k}{5}} f_{\xi_k}(\xi_k) A(a_{i(j)}(\xi_k),1) \Pi_{i,j,k}(\xi_k) d\xi_k, i(j) = 1 \end{cases} \quad (29)$$

If *m*=3 the *k*[th] integral in eq. (29) can be evaluated as [$A(b_k(\xi_1),-2)-A(a_k(\xi_1),-2)$].

Concluding the 2-way *SHO* subsection the $E[X_iY_j]$ term is expressed below:

$$E\left[X_iY_j\right] = \frac{C_{1,i}C_{k,j}}{P(\Omega_{1k}^m)} \times \begin{cases} \int_{-\infty}^{+\infty} 10^{-\frac{b\xi_1}{10}} f_{\xi_1}(\xi_1)d\xi_1 \int_{b_k(\xi_1)}^{a_k(\xi_1)} 10^{-\frac{b\xi_k}{10}} f_{\xi_k}(\xi_k)A(a_i(\xi_k),1)A(a_j(\xi_k),1)\Pi_{i,j,k}(\xi_k)d\xi_k, \begin{cases} i \neq k, \\ j \neq 1 \end{cases} \\ \int_{-\infty}^{+\infty} 10^{-\frac{b\xi_1}{10}} f_{\xi_1}(\xi_1)d\xi_1 \int_{b_k(\xi_1)}^{a_k(\xi_1)} f_{\xi_k}(\xi_k)A(a_j(\xi_k),1)\Pi_{i,j,k}(\xi_k)d\xi_k, \{i = k, j \neq 1\} \\ \int_{-\infty}^{+\infty} f_{\xi_1}(\xi_1)d\xi_1 \int_{b_k(\xi_1)}^{a_k(\xi_1)} 10^{-\frac{b\xi_k}{10}} f_{\xi_k}(\xi_k)A(a_i(\xi_k),1)\Pi_{i,j,k}(\xi_k)d\xi_k, \{i \neq k, j = 1\} \\ \int_{-\infty}^{+\infty} f_{\xi_1}(\xi_1)d\xi_1 \int_{b_k(\xi_1)}^{a_k(\xi_1)} f_{\xi_k}(\xi_k)\Pi_{i,j,k}(\xi_k)d\xi_k, \{i = k, j = 1\} \end{cases} \quad (30)$$

If $i=j\neq 1$ and $k$, $A(a_i(\xi_k),1)A(a_j(\xi_k),1)$ product should be replaced by $A(a_i(\xi_k),2)$. Finally, if $m=3$ and $i\neq k$ ($i=k$) the $k^{th}$ integral in eq. (30) is evaluated as $[A(b_k(\xi_1),-1)-A(a_k(\xi_1),-1)]$ ($[A(b_k(\xi_1),0)-A(a_k(\xi_1),0)]$).

*C) 3-way SHO Calculations*

As it was discussed in the 2-way *SHO* case the first and the second moment of $\beta_{1kl}$, can be approximated through a Taylor expansion of eq. (4). If we omit Taylor series terms higher than the second order the following expressions can be derived for the first and the second moment of $\beta_{1kl}$ (see Appendix II):

$$E[\beta_{1kl} \mid \underline{\xi} \in \Omega_{1kl}^m] = C_t \left\{ \frac{\overline{XYZ}}{\overline{XY}+\overline{XZ}+\overline{YZ}} - \frac{2\left[\overline{X^2}-(\overline{X})^2\right]\left[(\overline{Y}+\overline{Z})(\overline{YZ})^2\right]}{\left[\overline{XY}+\overline{XZ}+\overline{YZ}\right]^3} - \right.$$

$$-\frac{2\left[\overline{Y^2}-(\overline{Y})^2\right]\left[(\overline{X}+\overline{Z})(\overline{XZ})^2\right]}{\left[\overline{XY}+\overline{XZ}+\overline{YZ}\right]^3} - \frac{2\left[\overline{Z^2}-(\overline{Z})^2\right]\left[(\overline{X}+\overline{Y})(\overline{XY})^2\right]}{\left[\overline{XY}+\overline{XZ}+\overline{YZ}\right]^3} + \frac{2\left[\overline{XY}-\overline{X}\overline{Y}\right]\overline{XY}(\overline{Z})^3}{\left[\overline{XY}+\overline{XZ}+\overline{YZ}\right]^3} +$$

$$\left. +\frac{2\left[\overline{XZ}-\overline{X}\overline{Z}\right]\overline{XZ}(\overline{Y})^3}{\left[\overline{XY}+\overline{XZ}+\overline{YZ}\right]^3} + \frac{2\left[\overline{YZ}-\overline{Y}\overline{Z}\right]\overline{YZ}(\overline{X})^3}{\left[\overline{XY}+\overline{XZ}+\overline{YZ}\right]^3} \right\} \quad (31)$$

$$E[\beta_{1kl}^2 \mid \underline{\xi} \in \Omega_{1kl}^m] = C_t^2 \left\{ \left(\frac{\overline{XYZ}}{\overline{XY}+\overline{XZ}+\overline{YZ}}\right)^2 + \frac{\left[\overline{X^2}-(\overline{X})^2\right](\overline{YZ})^4}{\left[\overline{XY}+\overline{XZ}+\overline{YZ}\right]^4} + \frac{\left[\overline{Y^2}-(\overline{Y})^2\right](\overline{XZ})^4}{\left[\overline{XY}+\overline{XZ}+\overline{YZ}\right]^4} + \right.$$

$$+\frac{\left[\overline{Z^2}-(\overline{Z})^2\right](\overline{XY})^4}{\left[\overline{XY}+\overline{XZ}+\overline{YZ}\right]^4} + \frac{2\left[\overline{XY}-\overline{X}\overline{Y}\right](\overline{XY})^2(\overline{Z})^4}{\left[\overline{XY}+\overline{XZ}+\overline{YZ}\right]^4} + \frac{2\left[\overline{XZ}-\overline{X}\overline{Z}\right](\overline{XZ})^2(\overline{Y})^4}{\left[\overline{XY}+\overline{XZ}+\overline{YZ}\right]^4} +$$

$$\left. +\frac{2\left[\overline{YZ}-\overline{Y}\overline{Z}\right](\overline{YZ})^2(\overline{X})^4}{\left[\overline{XY}+\overline{XZ}+\overline{YZ}\right]^4} + \frac{2\overline{XYZ}}{\overline{XY}+\overline{XZ}+\overline{YZ}} \left[\frac{E[\beta_{1kl} \mid \underline{\xi} \in \Omega_{1kl}^m]}{C_t} - \frac{\overline{XYZ}}{\overline{XY}+\overline{XZ}+\overline{YZ}}\right] \right\} \quad (32)$$

where $\overline{X}$, $\overline{Y}$, $\overline{Z}$, $\overline{X^2}$, $\overline{Y^2}$, $\overline{Z^2}$, $\overline{XY}$, $\overline{XZ}$ and $\overline{YZ}$ correspond to $E[X \mid \underline{\xi} \in \Omega_{1kl}^m]$, $E[Y \mid \underline{\xi} \in \Omega_{1kl}^m]$, $E[Z \mid \underline{\xi} \in \Omega_{1kl}^m]$, $E[X^2 \mid \underline{\xi} \in \Omega_{1kl}^m]$, $E[Y^2 \mid \underline{\xi} \in \Omega_{1kl}^m]$, $E[Z^2 \mid \underline{\xi} \in \Omega_{1kl}^m]$, $E[XY \mid \underline{\xi} \in \Omega_{1kl}^m]$, $E[XZ \mid \underline{\xi} \in \Omega_{1kl}^m]$ and $E[YZ \mid \underline{\xi} \in \Omega_{1kl}^m]$, respectively.

Following a similar calculation scheme as in previous sections, the above mentioned E[.] terms can be expressed as a summation of all possible combinations of $E[X_i]$, $E[X_i^2]$, $E[X_i X_j]$, $E[Y_i]$, $E[Y_i^2]$, $E[Y_i Y_j]$, $E[Z_i]$, $E[Z_i^2]$, $E[Z_i Z_j]$, $E[X_i Y_j]$, $E[X_i Z_j]$ and $E[Y_i Z_j]$ terms.

In details, $E[X_i]$ and $E[Y_i]$ terms are given by the following equations:

$$E[X_i] = \frac{C_{1,i}}{P(\Omega_{1kl}^m)} \int_{-\infty}^{+\infty} 10^{-\frac{b\xi_1}{10}} f_{\xi_1}(\xi_1) d\xi_1 \times \begin{cases} \int_{b_k(\xi_1)}^{a_k(\xi_1)} f_{\xi_k}(\xi_k) d\xi_k \int_{b_l(\xi_k)}^{a_l(\xi_k)} f_{\xi_l}(\xi_l) A(a_i(\xi_l),1) \Pi_{i,k,l}(\xi_1) d\xi_l, i \neq k,l \\ \int_{b_k(\xi_1)}^{a_k(\xi_1)} 10^{\frac{b\xi_k}{10}} f_{\xi_k}(\xi_k) d\xi_k \int_{b_l(\xi_k)}^{a_l(\xi_k)} f_{\xi_l}(\xi_l) \Pi_{k,l}(\xi_1) d\xi_l, i = k \\ \int_{b_k(\xi_1)}^{a_k(\xi_1)} f_{\xi_k}(\xi_k) d\xi_k \int_{b_l(\xi_k)}^{a_l(\xi_k)} 10^{\frac{b\xi_l}{10}} f_{\xi_l}(\xi_l) \Pi_{k,l}(\xi_1) d\xi_l, i = l \end{cases} \quad (33)$$

$$E[Y_i] = \frac{C_{k,i}}{P(\Omega_{1kl}^m)} \times \begin{cases} \int_{-\infty}^{+\infty} f_{\xi_1}(\xi_1) d\xi_1 \int_{b_k(\xi_1)}^{a_k(\xi_1)} 10^{-\frac{b\xi_k}{10}} f_{\xi_k}(\xi_k) d\xi_k \int_{b_l(\xi_k)}^{a_l(\xi_k)} f_{\xi_l}(\xi_l) A(a_i(\xi_l),1) \Pi_{i,k,l}(\xi_1) d\xi_l, i \neq 1,l \\ \int_{-\infty}^{+\infty} 10^{\frac{b\xi_1}{10}} f_{\xi_1}(\xi_1) d\xi_1 \int_{b_k(\xi_1)}^{a_k(\xi_1)} 10^{-\frac{b\xi_k}{10}} f_{\xi_k}(\xi_k) d\xi_k \int_{b_l(\xi_k)}^{a_l(\xi_k)} f_{\xi_l}(\xi_l) \Pi_{k,l}(\xi_1) d\xi_l, i = 1 \\ \int_{-\infty}^{+\infty} f_{\xi_1}(\xi_1) d\xi_1 \int_{b_k(\xi_1)}^{a_k(\xi_1)} 10^{-\frac{b\xi_k}{10}} f_{\xi_k}(\xi_k) d\xi_k \int_{b_l(\xi_k)}^{a_l(\xi_k)} 10^{\frac{b\xi_l}{10}} f_{\xi_l}(\xi_l) \Pi_{k,l}(\xi_1) d\xi_l, i = l \end{cases} \quad (34)$$

where the various integration limits in eqs. (33), (34) and throughout this subsection are the same as the ones described in eq. (19). Apparently, the $E[Z_i]$ expressions are similar to the ones in eq. (34):

$$E[Z_i] = \frac{C_{l,i}}{P(\Omega_{1kl}^m)} \times \begin{cases} \int_{-\infty}^{+\infty} f_{\xi_1}(\xi_1) d\xi_1 \int_{b_k(\xi_1)}^{a_k(\xi_1)} f_{\xi_k}(\xi_k) d\xi_k \int_{b_l(\xi_k)}^{a_l(\xi_k)} 10^{-\frac{b\xi_l}{10}} f_{\xi_l}(\xi_l) A(a_i(\xi_l),1) \Pi_{i,k,l}(\xi_1) d\xi_l, i \neq 1,k \\ \int_{-\infty}^{+\infty} 10^{\frac{b\xi_1}{10}} f_{\xi_1}(\xi_1) d\xi_1 \int_{b_k(\xi_1)}^{a_k(\xi_1)} f_{\xi_k}(\xi_k) d\xi_k \int_{b_l(\xi_k)}^{a_l(\xi_k)} 10^{-\frac{b\xi_l}{10}} f_{\xi_l}(\xi_l) \Pi_{k,l}(\xi_1) d\xi_l, i = 1 \\ \int_{-\infty}^{+\infty} f_{\xi_1}(\xi_1) d\xi_1 \int_{b_k(\xi_1)}^{a_k(\xi_1)} 10^{\frac{b\xi_k}{10}} f_{\xi_k}(\xi_k) d\xi_k \int_{b_l(\xi_k)}^{a_l(\xi_k)} 10^{-\frac{b\xi_l}{10}} f_{\xi_l}(\xi_l) \Pi_{k,l}(\xi_1) d\xi_l, i = k \end{cases} \quad (35)$$

$E[X_i^2]$, $E[Y_i^2]$ and $E[Z_i^2]$ expressions can be obtained from eqs. (33)-(35) if we replace $C_{1,i}$ with $C_{1,i}^2$, $C_{k,i}$, with $C_{k,i}^2$, $C_{l,i}$, with $C_{l,i}^2$, $A(a_i,1)$ with $A(a_i,2)$ and $10^{\pm b\xi_1/10}, 10^{\pm b\xi_k/10}, 10^{\pm b\xi_l/10}$ with $10^{\pm b\xi_1/5}, 10^{\pm b\xi_k/5}, 10^{\pm b\xi_l/5}$. As in previous cases, $E[X_0]=E[Y_0]=E[Z_0]=(1-u)/P(\Omega_{1kl}^m)$, $E[X_0^2]=E[Y_0^2]=E[Z_0^2]$ $(1-u)^2/P(\Omega_{1kl}^m)$.

The $E[X_iX_j]$, $E[Y_iY_j]$ and $E[Z_iZ_j]$ terms can be obtained from eq. (33), (34) and (35) if $10^{-b\xi_1/10}$, $10^{-b\xi_k/10}$ and $10^{-b\xi_l/10}$ is replaced by $10^{-b\xi_1/5}$, $10^{-b\xi_k/5}$ and $10^{-b\xi_l/5}$, respectively. In addition, if $I$ and $j \neq k$ and $l$ in eq. (33), $I$ and $j \neq 1$ and $l$ in eq. (34) and $I$ and $j \neq 1$ and $k$ in eq. (35), $A(a_i(\xi_l),1)\Pi_{i,k,l}(\xi_l)$ should be replaced by

$$P(\xi_l) = A(a_i(\xi_l),1)A(a_j(\xi_l),1) \prod_{n=2, n \neq i,j,k,l}^{19} A(a_n(\xi_l),0).$$ Furthermore, if $i=k$ or $l$ in eq. (33), $i=1$ or $l$ in eq. (34) and $i=1$ or

$k$ in eq. (35), then $\Pi_{k,l}(\xi_l)$ should be replaced by $A(a_j(\xi_l),1)\Pi_{j,k,l}(\xi_l)$. Finally, if $j$ takes the latter $I$ values, the same expressions still apply if we interchange $I$ with $j$.

The cross product terms $E[X_iY_j]$, $E[X_iZ_j]$ and $E[Y_iZ_j]$ are expressed below:

$$E[X_iY_j] = \frac{C_{1,i}C_{k,j}}{P(\Omega_{1kl}^m)} \int_{-\infty}^{+\infty} 10^{-\frac{b\xi_1}{10}} f_{\xi_1}(\xi_1)d\xi_1 \int_{b_k(\xi_1)}^{a_k(\xi_1)} 10^{-\frac{b\xi_k}{10}} f_{\xi_k}(\xi_k)d\xi_k \int_{b_l(\xi_k)}^{a_l(\xi_k)} f_{\xi_l}(\xi_l)P(\xi_l)d\xi_l \quad (36)$$

$$E[X_iZ_j] = \frac{C_{1,i}C_{l,j}}{P(\Omega_{1kl}^m)} \int_{-\infty}^{+\infty} 10^{-\frac{b\xi_1}{10}} f_{\xi_1}(\xi_1)d\xi_1 \int_{b_k(\xi_1)}^{a_k(\xi_1)} f_{\xi_k}(\xi_k)d\xi_k \int_{b_l(\xi_k)}^{a_l(\xi_k)} 10^{-\frac{b\xi_l}{10}} f_{\xi_l}(\xi_l)P(\xi_l)d\xi_l \quad (37)$$

$$E[Y_iZ_j] = \frac{C_{k,i}C_{l,j}}{P(\Omega_{1kl}^m)} \int_{-\infty}^{+\infty} f_{\xi_1}(\xi_1)d\xi_1 \int_{b_k(\xi_1)}^{a_k(\xi_1)} 10^{-\frac{b\xi_k}{10}} f_{\xi_k}(\xi_k)d\xi_k \int_{b_l(\xi_k)}^{a_l(\xi_k)} 10^{-\frac{b\xi_l}{10}} f_{\xi_l}(\xi_l)P(\xi_l)d\xi_l \quad (38)$$

where we assume $i \neq k, l, j \neq 1, l$ and $i \neq j$ in eq. (36), $i \neq k, l, j \neq 1, k$ and $i \neq j$ in eq. (37) and $i \neq 1, l, j \neq 1, k$ and $i \neq j$ in eq. (38). If $i=k$ or $l$ the $10^{b\xi_i/10}$ term is transferred to the $k^{th}$ integral (thus $10^{-b\xi_k/10}$ vanishes in eq. (36)) or to the $l^{th}$ integral (thus $10^{-b\xi_l/10}$ vanishes in eq. (37)-(38)). In addition, if $i=1$ in eq. (38) $10^{b\xi_i/10}$ term is transferred to the $1^{st}$ integral. In all aforementioned cases $P(\xi_l)$ converts to $A(a_j(\xi_l),1)\Pi_{j,k,l}(\xi_l)$. If $j=1$ or $l$ the $10^{b\xi_j/10}$ term is transferred to the $1^{st}$ integral (thus $10^{-b\xi_1/10}$ vanishes in eqs. (36)-(37)) or to the $l^{th}$ integral. In addition, if $j=k$ in eqs. (37)-(38) the $10^{b\xi_j/10}$ term is transferred to the $k^{th}$ integral (thus $10^{-b\xi_k/10}$ vanishes in eq. (38)). In all aforementioned cases $P(\xi_l)$ converts to $A(a_i(\xi_l),1)\Pi_{i,k,l}(\xi_l)$. Finally, if $i=j \neq l$ in eq. (36), $i=j \neq k$ in eq. (37) and $i=j \neq 1$ in eq. (38) then $P(\xi_l)$ converts to $A(a_j(\xi_l),2)\Pi_{j,k,l}(\xi_l)$. Otherwise, if $i=j=l$ in eq. (36), $i=j=k$ in eq. (37) and $i=j=1$ in eq. (38) then the $10^{b\xi_l/5}$, $10^{b\xi_k/5}$ and $10^{b\xi_1/5}$ term appears in the $l^{th}$, $k^{th}$ and $1^{st}$ integral, respectively, whereas $P(\xi_l)$ converts to $\Pi_{k,l}(\xi_l)$.

## V. NUMERICAL RESULTS & DISCUSSION

First, a comparison between the calculations of the proposed theoretical model and the corresponding ones from an independent numerical simulation will be discussed. The calculations have been performed with respect to the expected value $E[\beta/\underline{\xi} \in \Omega^m]$ $(=\overline{\beta})$ and the standard deviation $\sqrt{E[\beta^2 | \underline{\xi} \in \Omega^m] - \left(E[\beta | \underline{\xi} \in \Omega^m]\right)^2}$ $(=\sigma_\beta)$ of power consumption. The under examination scenarios include different *MS* positions $(r,\theta)$, various path loss factors ($\alpha$) and standard deviations of shadowing losses ($\sigma$), as well as different *AS* sizes, cell selection thresholds (*cst*) and *SHO* thresholds (*sht*). The service parameters correspond to a typical voice service in WCDMA UMTS networks: $v=0.5$, $R=12.2$ Kbps, $W=3.84$ Mchips/s and $[E_b/I_o]_t=4.4$ dB. Finally, the orthogonality factor is $u=0.9$.

The numerical simulation model has been configured to generate 100.000 random shadowing samples according to a log-normal pdf. For each sample the cell selection and the handoff inequalities of Section III are examined, first to decide whether the sample refers to the cell under examination or not and next to decide which of the three handoff conditions is fulfilled. According to the latter criterion a power consumption sample is calculated using one of the equations (2)-(4), and next $\overline{\beta}$ and $\sigma_\beta$ is estimated using equations (13) and (14), respectively. In order to facilitate a tabulated comparison between the numerical results and the corresponding theoretical ones the results from 5 rounds of simulation runs have been averaged and presented in Tables I, II and III. Each Table refers to a different scenario and proves that theoretical and numerical estimations converge, which in turn proves the efficiency of the Taylor series approximation.

Next, in order to demonstrate the potential benefits from the adaptation of the proposed theoretical model the power consumption statistics will be further investigated. The under examination numerical results are illustrated in Figs. 3-8. Figs. 3, 5 and 7 depict $\overline{\beta}$ for *AS*=1, 2 and 3, respectively, versus the normalized distance $r_1/R_{\max}$. Figs 4, 6 and 8 depict $\sigma_\beta$ for the former scenarios.

Fig. 3 corresponds to a *HHO* scenario. According to the illustrated data $\overline{\beta}$ tends to increase, as expected, when the MS approaches the cell border. Near *BS* and up to a distance, $\overline{\beta}$ increases, when $\alpha$ and $\sigma$ take higher values. Nevertheless, this is not valid, when the *MS* approaches the cell border. Actually, close to the border a hostile propagation environment (i.e. high $\alpha$ and $\sigma$ values) results to less power consumption. This behavior can be explained, if we take into the account the possibility of handoff. Close to the border the *MS* tends to camp to another cell instead of sustaining the degradation of a hostile environment. Actually, this is more evident, when the cell selection criterion is more tight, i.e. *cst*=1 instead of *cst*=3, and camping to another cell is encouraged. The comments from Fig. 4 are rather similar to the ones in Fig. 3. The higher (lower) $\sigma_\beta$ appears, when $\alpha$ and $\sigma$ take lower (higher) and the cell selection algorithm decision criteria are relatively loose (tight). According to the aforementioned comments the cell selection imperfections burdens the system, when the propagation conditions are relatively good and *AS*=1. In such cases, the *MS* should be encouraged to camp to a neighbor cell.

Fig. 5 illustrates the expected value of power consumption, when *AS*=2 and thus a 2-way *SHO* is also possible. Fig. 5 also includes results for *AS*=1 for comparison reasons. According to the illustrated data the highest values of $\overline{\beta}$ appear, when $\alpha$ and $\sigma$ take low values as it was already mentioned in Fig.3. If we compare *AS*=1 and *AS*=2 results, it appears that the choice of *AS*=2 and more than this the encouragement of *SHO* is beneficial and this is more evident when the *MS* approaches the cell border. Actually, when *α* and *σ* take low values and the *MS* moves towards the cell border/corner *SHO* takes advantage of the good propagation conditions and allows one neighbor *BS* to participate instead of being a strong interferer. Fig. 6 illustrates $\sigma_\beta$ numerical results for the network scenarios examined in Fig. 5. According to the illustrated results, the option and more than this the encouragement of *SHO* reduces significantly $\sigma_\beta$ at least when compared to *AS*=1 scenarios. Concluding, the inclusion of a *SHO* option by setting *AS*=2, provides significant benefits, in terms of reducing $\bar{\beta}$ and $\sigma_\beta$, even in cases where the MS is located relatively close to the BS.

Fig. 7 illustrates the expected value of power consumption, when *AS*=3 and thus a 3-way *SHO* is also possible. According to the illustrated data the *AS*=3 choice gives slightly better results, when is compared with the relevant results of Fig. 5 and particular with the case of *σ*=8 dB. However, the encouragement of *SHO* (*sht*=3 dB) provides a significant reduction, when compared with the *AS*=1 and *AS*=2 choice and the case of *σ*=10 dB. Fig. 8 illustrates $\sigma_\beta$ for the network scenarios examined in Fig. 7. According to the illustrated results and the comparison with the relevant results in Fig.6, the choice of *AS*=3 and the encouragement of *SHO* provides a significant reduction of $\sigma_\beta$ and a location insensitive behavior.

Concluding the discussion on the aforementioned results it is worthwhile to mention that as it has been found in similar research works the resource allocation on CDMA networks strongly depends on the propagation conditions, the MS location and the various Radio Resource Management (RRM) settings. Thus, an optimized network performance definitely requires a cross layer approach and prediction models that can incorporate both physical layer and RRM parameters.

## VI. CONCLUSION

A theoretical statistical model that provides an estimation of the expected and standard deviation value of power consumption in the downlink direction has been developed for cellular CDMA networks. The proposed model supports the aforementioned calculations taking into account cell selection and handoff settings. In this context, present work contributes to a cross-layer approach, by establishing a theoretical framework, which facilitates performance evaluation and optimization of CDMA networks under specific radio propagation conditions as well as RRM settings. Current work can be extended with future studies in several directions. The most challenging future extension is to provide a joint pdf for power consumption based on the capability to estimate power consumption moments. Furthermore, present work provides estimations on a link level and thus an extension of the model in order to support performance evaluation on a network level is also another interesting research direction. A cross layer design approach aiming to develop an optimized soft handoff algorithm, which will take into account the proposed model's estimations, is another one possible future research topic. Finally, the under consideration numerical results are based on several assumptions, which can be easily rearranged. For example, it would be interesting to produce numerical results by taking into account unbalanced power allocation schemes among the *SHO* links or unequal traffic loads per cell.

# APPENDIX I

In the case of 2-way SHO connections, $\beta_{1k}$ power consumption metric can be expressed in the form of the following function:

$$\beta(X,Y) = \left(\frac{1}{X} + \frac{1}{Y}\right)^{-1} = \frac{XY}{X+Y} \qquad (I.1)$$

Using a Taylor expansion in the neighborhood of $E[X(\xi)] = \overline{X}$ and $E[Y(\xi)] = \overline{Y}$, where Taylor series terms higher than the second order are omitted, and next taking the average value of this expression we conclude after a few straightforward calculations to:

$$E[\beta(X,Y)] = \beta(\overline{X},\overline{Y}) + E[(X-\overline{X})^2]\frac{\partial^2}{\partial X^2}[\beta(X,Y)]_{\substack{X=\overline{X}\\Y=\overline{Y}}} + E[(Y-\overline{Y})^2]\frac{\partial^2}{\partial Y^2}[\beta(X,Y)]_{\substack{X=\overline{X}\\Y=\overline{Y}}} +$$

$$+ 2E[(X-\overline{X})(Y-\overline{Y})]\frac{\partial^2}{\partial X \partial Y}[\beta(X,Y)]_{\substack{X=\overline{X}\\Y=\overline{Y}}} \qquad (I.2)$$

where

$$\frac{\partial^2}{\partial X^2}[\beta(X,Y)]_{\substack{X=\overline{X}\\Y=\overline{Y}}} = -\frac{2(\overline{Y})^2}{(\overline{X}+\overline{Y})^3}, \frac{\partial^2}{\partial Y^2}[\beta(X,Y)]_{\substack{X=\overline{X}\\Y=\overline{Y}}} = -\frac{2(\overline{X})^2}{(\overline{X}+\overline{Y})^3}, \frac{\partial^2}{\partial X \partial Y}[\beta(X,Y)]_{\substack{X=\overline{X}\\Y=\overline{Y}}} = \frac{2\overline{XY}}{(\overline{X}+\overline{Y})^3}$$

(I.3)

By taking the square power of the above mentioned Taylor series expansion and omitting higher order terms, we conclude, after some manipulation, to the following expression regarding $E[\beta^2(X,Y)]$:

$$E[\beta^2(X,Y)] = \beta^2(\overline{X},\overline{Y}) + E[(X-\overline{X})^2]\left(\frac{\partial}{\partial X}[\beta(X,Y)]_{\substack{X=\overline{X}\\Y=\overline{Y}}}\right)^2 + E[(Y-\overline{Y})^2]\left(\frac{\partial}{\partial Y}[\beta(X,Y)]_{\substack{X=\overline{X}\\Y=\overline{Y}}}\right)^2 +$$

$$+ 2E[(X-\overline{X})(Y-\overline{Y})]\left(\frac{\partial}{\partial X}[\beta(X,Y)]_{\substack{X=\overline{X}\\Y=\overline{Y}}}\right)\left(\frac{\partial}{\partial Y}[\beta(X,Y)]_{\substack{X=\overline{X}\\Y=\overline{Y}}}\right) + 2\beta(\overline{X},\overline{Y})\{E[\beta(X,Y)] - \beta(\overline{X},\overline{Y})\}$$

(I.4)

where

$$\frac{\partial}{\partial X}[\beta(X,Y)]_{\substack{X=\overline{X}\\Y=\overline{Y}}} = \frac{(\overline{Y})^2}{(\overline{X}+\overline{Y})^2}, \frac{\partial}{\partial Y}[\beta(X,Y)]_{\substack{X=\overline{X}\\Y=\overline{Y}}} = \frac{(\overline{X})^2}{(\overline{X}+\overline{Y})^2} \qquad (I.5)$$

# APPENDIX II

In the case of 3-way SHO connections, $\beta_{1kl}$ power consumption metric can be expressed in the form of the following function:

$$\beta(X,Y,Z) = \left(\frac{1}{X} + \frac{1}{Y} + \frac{1}{Z}\right)^{-1} = \frac{XYZ}{XY + XZ + YZ} \qquad (II.1)$$

Using a Taylor expansion as in Appendix II, we conclude after a few straightforward calculations to:

$$E[\beta(X,Y,Z)] = \beta(\overline{X},\overline{Y},\overline{Z}) + E\left[(X-\overline{X})^2\right]\frac{\partial^2}{\partial X^2}[\beta(X,Y,Z)]\Big|_{\substack{X=\overline{X}\\Y=\overline{Y}\\Z=\overline{Z}}} + E\left[(Y-\overline{Y})^2\right]\frac{\partial^2}{\partial Y^2}[\beta(X,Y,Z)]\Big|_{\substack{X=\overline{X}\\Y=\overline{Y}\\Z=\overline{Z}}} +$$

$$+ E\left[(Z-\overline{Z})^2\right]\frac{\partial^2}{\partial Z^2}[\beta(X,Y,Z)]\Big|_{\substack{X=\overline{X}\\Y=\overline{Y}\\Z=\overline{Z}}} + 2E\left[(X-\overline{X})(Y-\overline{Y})\right]\frac{\partial^2}{\partial X\partial Y}[\beta(X,Y,Z)]\Big|_{\substack{X=\overline{X}\\Y=\overline{Y}\\Z=\overline{Z}}} +$$

$$+ 2E\left[(X-\overline{X})(Z-\overline{Z})\right]\frac{\partial^2}{\partial X\partial Z}[\beta(X,Y,Z)]\Big|_{\substack{X=\overline{X}\\Y=\overline{Y}\\Z=\overline{Z}}} + 2E\left[(Y-\overline{Y})(Z-\overline{Z})\right]\frac{\partial^2}{\partial Y\partial Z}[\beta(X,Y,Z)]\Big|_{\substack{X=\overline{X}\\Y=\overline{Y}\\Z=\overline{Z}}} \quad (II.2)$$

where

$$\frac{\partial^2}{\partial X^2}[\beta(X,Y,Z)]\Big|_{\substack{X=\overline{X}\\Y=\overline{Y}\\Z=\overline{Z}}} = -\frac{2(\overline{Y}\overline{Z})^2(\overline{Y}+\overline{Z})}{(\overline{X}\overline{Y}+\overline{X}\overline{Z}+\overline{Y}\overline{Z})^3}, \frac{\partial^2}{\partial Y^2}[\beta(X,Y,Z)]\Big|_{\substack{X=\overline{X}\\Y=\overline{Y}\\Z=\overline{Z}}} = -\frac{2(\overline{X}\overline{Z})^2(\overline{X}+\overline{Z})}{(\overline{X}\overline{Y}+\overline{X}\overline{Z}+\overline{Y}\overline{Z})^3},$$

$$\frac{\partial^2}{\partial Z^2}[\beta(X,Y,Z)]\Big|_{\substack{X=\overline{X}\\Y=\overline{Y}\\Z=\overline{Z}}} = -\frac{2(\overline{X}\overline{Y})^2(\overline{X}+\overline{Y})}{(\overline{X}\overline{Y}+\overline{X}\overline{Z}+\overline{Y}\overline{Z})^3} \qquad (II.3)$$

$$\frac{\partial^2}{\partial X\partial Y}[\beta(X,Y,Z)]\Big|_{\substack{X=\overline{X}\\Y=\overline{Y}\\Z=\overline{Z}}} = \frac{2\overline{X}\overline{Y}(\overline{Z})^3}{(\overline{X}\overline{Y}+\overline{X}\overline{Z}+\overline{Y}\overline{Z})^3}, \frac{\partial^2}{\partial X\partial Z}[\beta(X,Y,Z)]\Big|_{\substack{X=\overline{X}\\Y=\overline{Y}\\Z=\overline{Z}}} = \frac{2\overline{X}(\overline{Y})^3\overline{Z}}{(\overline{X}\overline{Y}+\overline{X}\overline{Z}+\overline{Y}\overline{Z})^3}$$

$$\frac{\partial^2}{\partial Y\partial Z}[\beta(X,Y,Z)]\Big|_{\substack{X=\overline{X}\\Y=\overline{Y}\\Z=\overline{Z}}} = \frac{2(\overline{X})^3\overline{Y}\overline{Z}}{(\overline{X}\overline{Y}+\overline{X}\overline{Z}+\overline{Y}\overline{Z})^3} \qquad (II.4)$$

Using the square power of the Taylor series expansions and omitting higher order terms, we conclude to the following expression regarding $E[\beta^2(X,Y,Z)]$:

$$E[\beta^2(X,Y,Z)] = \beta^2(\overline{X},\overline{Y},\overline{Z}) + E\left[(X-\overline{X})^2\right]\left(\frac{\partial}{\partial X}[\beta(X,Y,Z)]\Big|_{\substack{X=\overline{X}\\Y=\overline{Y}\\Z=\overline{Z}}}\right)^2 +$$

$$+ E\left[(Y-\overline{Y})^2\right]\left(\frac{\partial}{\partial Y}[\beta(X,Y,Z)]\Big|_{\substack{X=\overline{X}\\Y=\overline{Y}\\Z=\overline{Z}}}\right)^2 + E\left[(Z-\overline{Z})^2\right]\left(\frac{\partial}{\partial Z}[\beta(X,Y,Z)]\Big|_{\substack{X=\overline{X}\\Y=\overline{Y}\\Z=\overline{Z}}}\right)^2 +$$

$$+ 2E\left[(X-\overline{X})(Y-\overline{Y})\right]\left(\frac{\partial}{\partial X}[\beta(X,Y,Z)]\Big|_{\substack{X=\overline{X}\\Y=\overline{Y}\\Z=\overline{Z}}}\right)\left(\frac{\partial}{\partial Y}[\beta(X,Y,Z)]\Big|_{\substack{X=\overline{X}\\Y=\overline{Y}\\Z=\overline{Z}}}\right) +$$

$$+ 2E\left[(X-\overline{X})(Z-\overline{Z})\right]\left(\frac{\partial}{\partial X}[\beta(X,Y,Z)]_{\substack{X=\overline{X}\\Y=\overline{Y}\\Z=\overline{Z}}}\right)\left(\frac{\partial}{\partial Z}[\beta(X,Y,Z)]_{\substack{X=\overline{X}\\Y=\overline{Y}\\Z=\overline{Z}}}\right)+$$

$$+ 2E\left[(Y-\overline{Y})(Z-\overline{Z})\right]\left(\frac{\partial}{\partial Y}[\beta(X,Y,Z)]_{\substack{X=\overline{X}\\Y=\overline{Y}\\Z=\overline{Z}}}\right)\left(\frac{\partial}{\partial Z}[\beta(X,Y,Z)]_{\substack{X=\overline{X}\\Y=\overline{Y}\\Z=\overline{Z}}}\right)+$$

$$+ 2\beta(\overline{X},\overline{Y},\overline{Z})\left\{E[\beta(X,Y,Z)] - \beta(\overline{X},\overline{Y},\overline{Z})\right\} \tag{II.5}$$

where

$$\frac{\partial}{\partial X}[\beta(X,Y,Z)]_{\substack{X=\overline{X}\\Y=\overline{Y}\\Z=\overline{Z}}} = \frac{(\overline{Y}\overline{Z})^2}{(\overline{X}\overline{Y}+\overline{X}\overline{Z}+\overline{Y}\overline{Z})^2}, \frac{\partial}{\partial Y}[\beta(X,Y,Z)]_{\substack{X=\overline{X}\\Y=\overline{Y}\\Z=\overline{Z}}} = \frac{(\overline{X}\overline{Z})^2}{(\overline{X}\overline{Y}+\overline{X}\overline{Z}+\overline{Y}\overline{Z})^2},$$

$$\frac{\partial}{\partial Z}[\beta(X,Y,Z)]_{\substack{X=\overline{X}\\Y=\overline{Y}\\Z=\overline{Z}}} = \frac{(\overline{X}\overline{Y})^2}{(\overline{X}\overline{Y}+\overline{X}\overline{Z}+\overline{Y}\overline{Z})^2} \tag{II.6}$$

**Table I. Theoretical vs Numerical Simulation Estimations for AS=1**

| AS=1<br>$\theta=15^o, a=3, \sigma=8, cst=1$ | | $r=0.6R_{max}$ | $r=0.7R_{max}$ | $r=0.8R_{max}$ | $r=0.9R_{max}$ | $r=1.0R_{max}$ |
|---|---|---|---|---|---|---|
| Theoretical Model | $\bar{\beta}$ | 0.0035164 | 0.0043695 | 0.0051464 | 0.0058282 | 0.0064095 |
| | $\sigma_\beta$ | 0.0032682 | 0.0037155 | 0.0040348 | 0.0042499 | 0.0043742 |
| Numerical Model | $\bar{\beta}$ | 0.0035124 | 0.0043708 | 0.0051283 | 0.0057727 | 0.0064118 |
| | $\sigma_\beta$ | 0.0032512 | 0.0036875 | 0.0040272 | 0.0041949 | 0.0043510 |

**Table II. Theoretical vs Numerical Simulation Estimations for AS=2**

| AS=2<br>$\theta=30^o, a=3, \sigma=8, cst=1, sht=3$ | | $r=0.6R_{max}$ | $r=0.7R_{max}$ | $r=0.8R_{max}$ | $r=0.9R_{max}$ | $r=1.0R_{max}$ |
|---|---|---|---|---|---|---|
| Theoretical Model | $\bar{\beta}$ | 0.0031774 | 0.0036789 | 0.0040796 | 0.0043946 | 0.0046417 |
| | $\sigma_\beta$ | 0.0017504 | 0.0018292 | 0.0018618 | 0.0018691 | 0.0018788 |
| Numerical Model | $\bar{\beta}$ | 0.0031854 | 0.0036904 | 0.0040845 | 0.0044148 | 0.0046605 |
| | $\sigma_\beta$ | 0.0017671 | 0.0018521 | 0.0018910 | 0.0019045 | 0.0019193 |

**Table III. Theoretical vs Numerical Simulation Estimations for AS=3**

| AS=3 $\theta=0^o$, $a=4, \sigma=10, cst=1$, $sht=3$ | | r=0.6$R_{max}$ | r=0.7$R_{max}$ | r=0.8$R_{max}$ | r=0.9$R_{max}$ | r=1.0$R_{max}$ |
|---|---|---|---|---|---|---|
| Theoretical Model | $\overline{\beta}$ | 0.0016295 | 0.0019904 | 0.0022905 | 0.0025476 | 0.0027399 |
| | $\sigma_\beta$ | 0.0010969 | 0.0012158 | 0.0012493 | 0.0012537 | 0.0012392 |
| Numerical Model | $\overline{\beta}$ | 0.0016487 | 0.0020319 | 0.0023516 | 0.0026368 | 0.0028613 |
| | $\sigma_\beta$ | 0.0011215 | 0.0012532 | 0.0013089 | 0.0013289 | 0.0013308 |

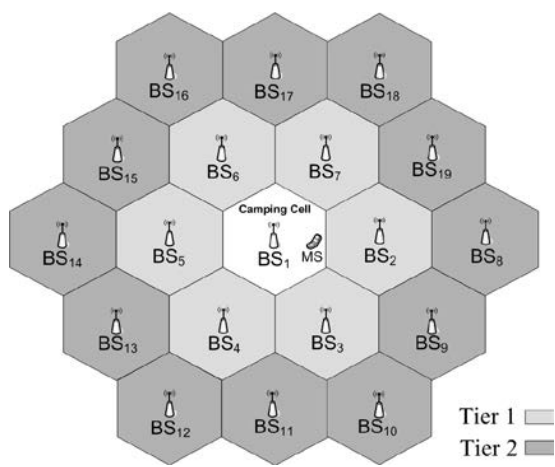

Fig. 1. The considered cellular network

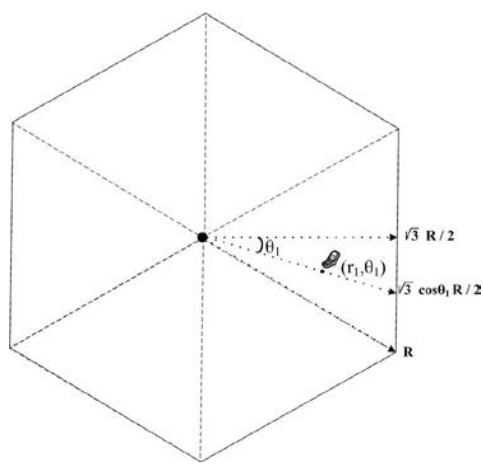

Fig. 2. The cell geometry and the spatial coordinates $r$, $\theta$.

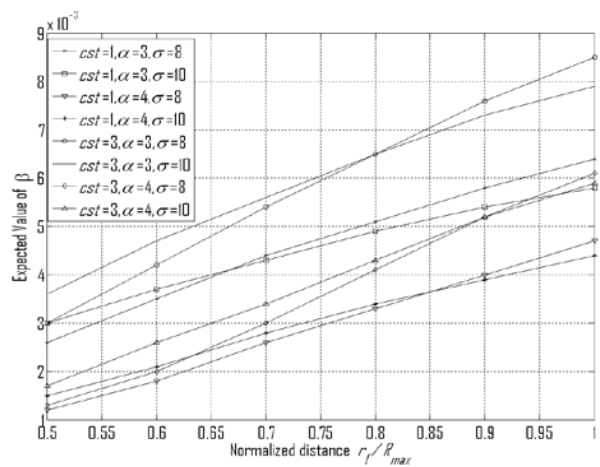

Fig 3. Expected value of power consumption versus normalized distance $r_1/R_{max}$ (AS=1 and $\theta=15^o$)

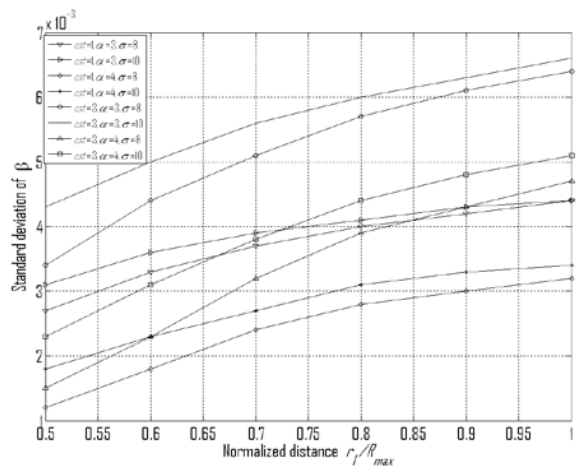

Fig 4. Standard deviation of power consumption versus normalized distance $r_1/R_{max}$ (AS=1 and $\theta=15^o$)

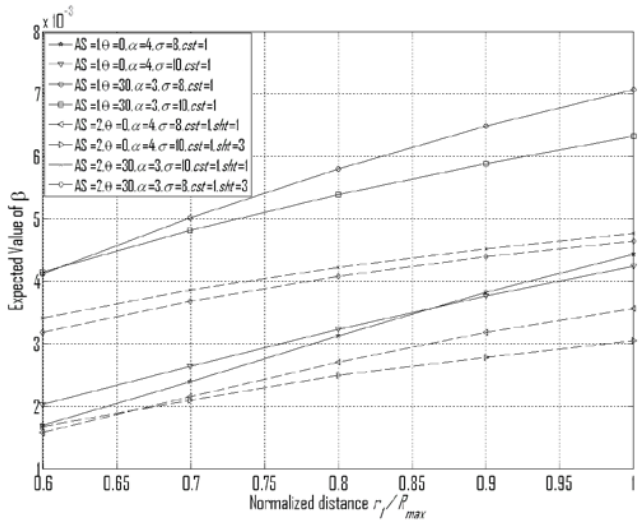

Fig 5. Expected value of power consumption versus normalized distance $r_1/R_{max}$ (AS=1, 2).

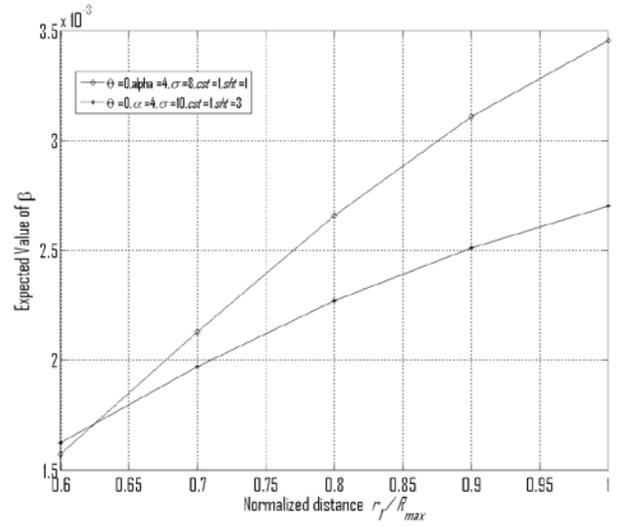

Fig 7. Expected value of power consumption versus normalized distance $r_1/R_{max}$ (AS=3).

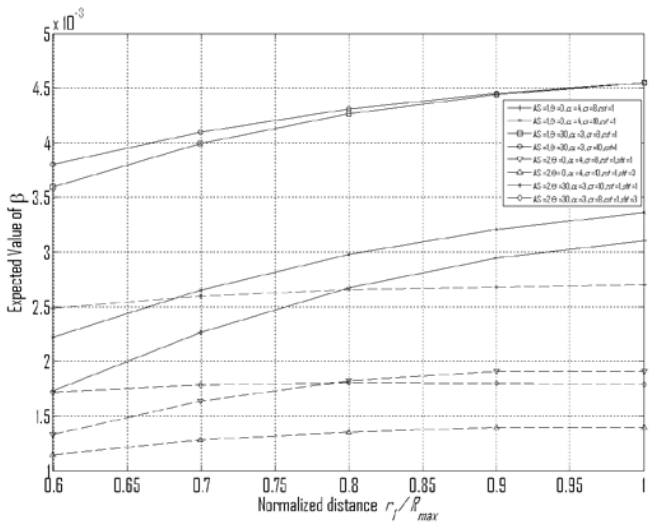

Fig 6. Standard deviation of power consumption versus normalized distance $r_1/R_{max}$ (AS=1, 2).

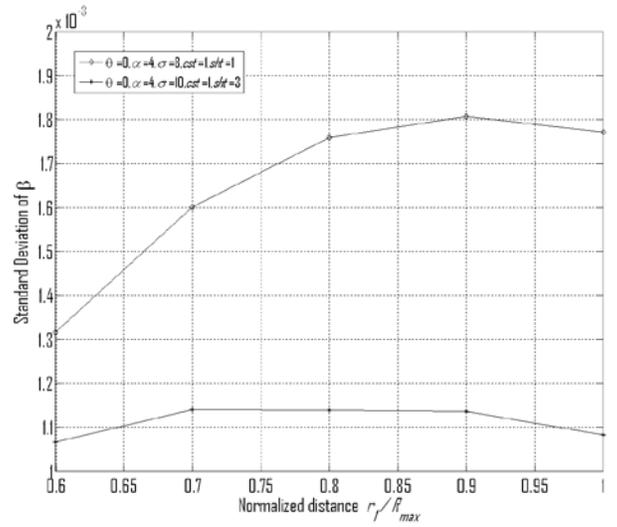

Fig 8. Standard deviation of power consumption versus normalized distance $r_1/R_{max}$ (AS=3).